\documentclass[reprint,amsmath,amssymb,aps,nofootinbib,superscriptaddress]{revtex4-1}

\usepackage{graphicx}
\usepackage{amsmath}
\usepackage{amssymb}
\usepackage{slashed,hyperref}
\def\be{\begin{equation}}
\def\ee{\end{equation}}
\def\bea{\begin{eqnarray}}
\def\eea{\end{eqnarray}}
\newcommand{ \bs } { \! \! }
\newcommand{\ba}{\begin{array}}
\newcommand{\ea}{\end{array}}
\newcommand{\ds}{\displaystyle}

\begin{document}
\title{Infinite-redshift localized states of Dirac fermions under Einsteinian gravity}

\author{Daniel \surname{Bakucz Can\'{a}rio}}
\altaffiliation[Current address:  ]{Max-Planck-Institut f{\"u}r Kernphysik, Saupfercheckweg 1, 69117 Heidelberg, Germany}
%\affiliation{SUPA, School of Physics \& Astronomy, University of St Andrews, North Haugh, St Andrews, Fife, KY16 9SS, UK}
\author{Sam Lloyd}
%\affiliation{SUPA, School of Physics \& Astronomy, University of St Andrews, North Haugh, St Andrews, Fife, KY16 9SS, UK}
\author{Keith Horne}
%\affiliation{SUPA, School of Physics \& Astronomy, University of St Andrews, North Haugh, St Andrews, Fife, KY16 9SS, UK}
\author{Chris A. Hooley}
\affiliation{SUPA, School of Physics \& Astronomy, University of St Andrews, North Haugh, St Andrews, Fife, KY16 9SS, UK}

\date{\today} 
\begin{abstract}
We present a set of localized states for an even number of Dirac fermions under Einsteinian gravity that have an infinite central redshift.  Near the center of the localized state the components of the Dirac spinor and the spacetime metric all show simple power-law dependences on the radial distance; further out the fermionic wave function decays to zero and the spacetime becomes asymptotically flat.  We show that this `central' solution of the equations of motion can be used to understand much of the structure observed by Finster, Smoller, and Yau [{\it Phys.\ Rev.\ D\/} {\bf 59}, 104020 (1999)] in their numerical solutions of the same problem at finite central redshift.
\end{abstract}
\maketitle
\section{Introduction}
Twenty-one years ago, in an important paper \cite{FSY1}, Finster, Smoller, and Yau (henceforth FSY) formulated the problem of two Dirac fermions coupled to Einsteinian gravity.  Their stated intention was to treat this as a model problem in which the interplay of the Heisenberg uncertainty principle and the mutual gravitational attraction of the fermions could be explored. This was followed by a string of papers that developed the theory by including electromagnetic effects \cite{FSY2}, angular momentum \cite{FSY3,FSY4}, and Yang-Mills interactions \cite{Bernard}. On the face of it, this approach has a clear deficiency: the gravitational sector is treated classically, an approximation whose conditions of validity are not obvious.  Nonetheless, one might hope that this kind of analysis can at least exhibit the basic phenomena to be expected when quantum particles bind gravitationally, and it is possible that phase space restrictions --- rather than a small parameter analogous to QED's fine structure constant --- might reduce the importance of quantum-gravitational corrections, at least for the low-lying states. Indeed, forty years prior, Brill and Wheeler \cite{BrillWheeler} had already considered a classical coupling of gravity and quantum mechanics in order to describe neutrinos, and later Lee and Pang \cite{Lee,LeePang} modeled fermion soliton stars, very similar to the FSY system, by again coupling general relativity to quantum mechanical quantities.

FSY's approach was to couple Einsteinian general relativity to Dirac's relativistic quantum mechanics. They focused on the problem of two identical Dirac fermions in a spin-singlet state, which allowed them to look for solutions in which both the fermion density and the spacetime metric were functions of only the radial coordinate, $r$.  They reduced this problem to a set of coupled ordinary differential equations for four fields:\ $\alpha(r)$ and $\beta(r)$, roughly the particle-like and hole-like parts of the Dirac spinor, and $A(r)$ and $T(r)$, roughly the length-contraction and time-dilation factors from the spacetime metric.  Importantly, they imposed particular conditions on their solutions to these equations:
\begin{enumerate}
\item[(i)]
that the two-fermion wave function be normalized;
\item[(ii)]
that the spacetime be asymptotically flat as $r \to \infty$;
\item[(iii)]
that all four fields behave regularly at $r=0$.
\end{enumerate}
As applied to $T(r)$, this last assumption amounts to requiring that the 
localised state's central redshift 
\be
z \equiv T(0)-1
\ee
be finite. FSY found distinct one-parameter families of solutions, one for the ground state with $n=0$ and one for each of the excited states with $n>0$ nodes in the fermion fields. The solutions within each family may be parameterized by their central redshift $z$, which sets the central value of the time-dilation field $T(0)=1+z$ at $r=0$. 
Along this sequence of solutions, the fermion mass $m$, fermion energy $\omega$, the ADM mass $M$, and the soliton size $R$ all depend on the central redshift $z$. 

 As $z$ increases, FSY's numerical solutions appear to `home in' on a particular solution with infinite redshift, and with $m$, $\omega$, $M$, and $R$ showing damped oscillations around well-defined asymptotic values.  The asymptotic value of the ADM mass, $M$, at high central redshift is greater than the rest masses of the two fermions, $2\,m$, meaning that high-redshift states are energetically unbound and unstable to fission; nonetheless, high-redshift states remain worthy of study both as a well defined limit of the FSY problem and as a resonance potentially observable in scattering experiments. In this paper, we analyze the FSY equations from this point of view. By exploring this special infinite red-shift solution, we can ``re-assemble'' the original FSY system and illuminate an underlying 4-zone structure of the solutions that was hitherto obscured by their numerical nature.

The paper is organised as follows. In Section~\ref{s:equations}, we recap the derivation of the FSY equations, generalizing from the two-fermion case to that of an arbitrary even integer number of fermions.  In Section~\ref{s:purepower}, we show analytically that, if we relax all three of FSY's boundary conditions and in addition set the fermion mass $m$ to zero, the FSY equations then admit a solution in which all fields are power-law functions of $r$.

In Section~\ref{s:asympflat}, we restore the fermion mass, which allows us to reinstate boundary conditions (i) and (ii).  We show that in this case one can still find numerically an infinite-redshift solution that resembles the pure power-law solution in an `inner zone' at small values of $r$, before crossing over into an `outer zone' at larger $r$ in which the fermion density decays exponentially and the spacetime metric tends to the Schwarzschild form.

In Section~\ref{s:wavezone}, we look at higher-lying excited-state solutions in which the fermionic fields have more zeros, and show that in this case there are three zones:\ a power-law zone resembling the analytic solution from section \ref{s:purepower}; a `wave zone' that looks approximately like non-relativistic particles trapped in a gravitational potential well; and an outer zone that resembles the one found in section~\ref{s:asympflat}.

In Section~\ref{s:fourzones}, we reintroduce condition (iii) --- regularity of the field $T(r)$ at the origin --- and show that this inserts a fourth `core' zone near the origin which matches on to the power-law zone at a value of $r$ that we estimate.  The core shrinks to zero in the high-redshift limit in which the FSY
solutions spiral around and converge onto our infinite-redshift solution.

In Section~\ref{s:poweroscillations}, we analyze the stability of the power-law solution to the perturbations caused by imperfect matching to the core-zone solution. We establish a strong link between these perturbations and the spiral characteristics of figures for the binding and gravitational energy in the original FSY paper.  In Section~\ref{s:physprop}, we discuss how the physical properties of the fermions differ between different zones of the solutions, concentrating on the fermion energy density and the radial and azimuthal pressures.  Finally, in Section~\ref{s:summary}, we summarize our results and discuss possible directions for further work.

 For convenience, we will sometimes refer to these `particle-like'  stationary localised states as `solitons',
as they are localised solutions of non-linear equations,
some of which are energetically bound relative to the
infinitely dispersed configuration of the constituent fermions.
Note, however, that time-dependent studies of their collisions
have not yet been performed.

\section{FSY's equations of motion for two-fermion gravitationally localised states}
\label{s:equations}

The physical system considered by Finster, Smoller, and Yau in their 1999 paper \cite{FSY1} is a pair of neutral fermions, each of mass $m$ and spin $j=1/2$, in a deformable spacetime governed by Einstein's general theory of relativity. 
 This two-fermion case can be easily generalized to the case of an arbitrary even integer number
$N=2\,j+1$ of fermions filling the shell of states with angular momentum $j$ 
to form a spherically symmetric singlet state.
The action for this system is
\begin{equation}  \label{eq:action}
	S = \int d^4 x \,\sqrt{-g} \,\left( \frac{R}{16\,\pi\, G} 
	+ \sum_{a=1}^{N} \overline{\psi}_a  \left( \slashed{D}- m \right) \psi_a \right).
\end{equation}
Here $g$ is the determinant of $g_{\mu \nu}$, the metric tensor of the deformable spacetime; $R \equiv R^{\mu}_{\mu}$ is the Ricci scalar; $G$ is the gravitational constant; $\psi_a$ is the Dirac spinor of fermion $a$; and
\be
\overline{\psi}_a = \psi_a^{\dagger}\, \gamma^0 = \psi_a^{\dagger} 
	\, \text{diag}\left( 1, 1, -1, -1 \right)
%\left( \begin{array}{cccc} \,1\,&\,0\,&0&0 \\ 0&1&0&0 \\ 0&0&-1&0 \\ 0&0&0&-1 \end{array} \right)
\ee
is the adjoint spinor.  $\slashed{D} \equiv i \,G^{\nu} \,D_{\nu}$ is the Dirac operator; in a curved space-time the Dirac matrices, $G^{\nu}$, satisfy the anticommutation relation \cite{Weinberg}
\be
\lbrace G^{\mu}, G^{\nu} \rbrace = 2 \, g^{\mu \nu} I,
\ee
where $I$ is the $4 \times 4$ identity matrix.

We follow FSY in assuming a static, spherically symmetric form for the metric tensor of the spacetime.
With spacetime coordinates $x^\mu=\left(t,r,\theta,\phi\right)$, the metric's covariant components
are
\begin{equation}
g_{\mu \nu} =\text{diag} \left(-T^{-2}, A^{-1}, r^2, r^2 \sin^2 \theta\right), \label{metric0}
\end{equation}
where $T(r)$ and $A(r)$ are functions of $r$ only.  
Here and henceforth we adopt units in which $\hbar = c = 1$.

In writing (\ref{eq:action}) we have assumed that the $N$ fermions are in a Slater-determinant (Fock) state in which each of the single-particle states in a given angular momentum shell is occupied precisely once.  
The FSY ansatz for the single-particle wave functions is given by
\bea
\Psi^{\pm}_{j \,k \,\omega} & = & e^{-i \,\omega \,t} \frac{\sqrt{T(r)}}{r} \left(\begin{matrix}
 \alpha^{\pm}_{j \, \omega}(r) \\[6pt]
i  \,\beta^{\pm}_{j \,  \omega}(r)\end{matrix}
\right) \, \chi^{k}_{j\mp1}(\theta,\phi)
\ ,
\eea
where $-j\le k \le j$ is the $z$ component of the angular momentum $j$,
$\pm$ is the parity (chirality) of the state.
The real functions $\alpha^\pm_{j\,\omega}(r)$ and 
$\beta^\pm_{j\,\omega}(r)$ define the radial structure of
the particle-like and hole-like components, respectively,
while the 2-spinors $\chi^{k}_{j \pm \frac{1}{2}}(\theta,\phi)$ give the angular dependence
of the wavefunction 
(an explicit expression for which can be found in Appendix~\ref{appendixA}). 
The general solution to the Dirac equation is then written as a linear combination of these wavefunctions. 

%\sout{Both $\Phi^{c}_{j k \omega 1}$ and $\Phi^{c}_{j k \omega 2}$ in  $\Psi^{c}_{j k \omega}$ are real\cite{FSY3} and are denoted by $\alpha$ and $\beta$ in analogy to the parameters of the previous EDM system.} \textbf{DBC: I suggest removing this sentence and directly introducing the reader to the $\alpha, \beta$ notation. Would like to double check with you that I'm not introducing a falsehood, though.}

Extremising the action $S$ under these assumptions about the Dirac and metric fields, we obtain the following set of four coupled ordinary differential equations for the fermion fields $\alpha$ and $\beta$ and the metric fields $A$ and $T$:
\begin{eqnarray}
\label{ed1}
	\sqrt{A} \,\frac{d\alpha}{dr} 
	& = & +\frac{\sigma\,N}{2\,r}\, \alpha - (E+m)\,\beta
\ , \\ \label{ed2}
	\sqrt{A} \,\frac{d\beta}{dr} 
	& = &   - \frac{\sigma\,N}{2\,r}\, \beta + (E-m)\, \alpha
\ , \\ \label{ed3}
	r \,\frac{dA}{dr} 
	& = & 1- A - 8\, \pi \,G\,T \,  N \, E \,(\alpha^2 + \beta^2)
\ , \\ 
	\frac{2\,r\,A}{T} \,\frac{dT}{dr} 
	&=&  A - 1 - 8 \,\pi \,  G  \, T  \, N \, \times
	 \nonumber 
\\ & & %\qquad \left. \displaystyle \phantom{ \displaystyle \frac{ \alpha}{r}}
\bs\bs\bs\bs\bs\bs\bs\bs\bs\bs\bs
	 \left( E \,  (\alpha^2 + \beta^2)  %\right.
	 - \sigma\, N \, \frac{\alpha \,\beta}{r} 
	 - m   \, (\alpha^2 - \beta^2)\right)
\ . \label{ed4}
\end{eqnarray}
Here $N=2\,j+1$ is the (even) number of fermions that occupy the filled shell
with angular momentum $j$,  $\sigma=+1$ for even and $-1$ for odd parity states,
and the `local fermion energy' $E(r)$ is defined as
\be
	E(r) \equiv \omega\, T(r)
\ .
\ee

Equations (\ref{ed1})-(\ref{ed4}) form a system of coupled first-order, non-linear ordinary differential equations in the two metric and two fermion fields, with parameters $N$, $\sigma$, $m$ and $\omega$. Such equations rarely have analytic, closed form solutions; nonetheless, we have identified a family of solutions for a reduced problem that displays exactly these qualities.

\section{A pure power-law solution of the FSY equations}
\label{s:purepower}
In the case where the fermion mass $m$ can be neglected (or indeed is actually zero), we obtain the following equations of motion:
\begin{eqnarray}
\label{ed1a}
	\sqrt{A} \,\frac{d\alpha}{dr} 
	&=& + \frac{ \sigma \, N } { 2 \, r } \, \alpha - \omega \,T \, \beta
\ , \\ \label{ed2a}
	\sqrt{ A } \,\frac{ d\beta } { dr } 
	&=&   - \frac{ \sigma \,N } { 2 \, r } \, \beta + \omega \, T \, \alpha
\ , \\ \label{ed3a}
	r \, \frac{dA}{dr} 
	&=& 1- A - 8 \, \pi \, G \, N \, \omega \, T^2 \, (\alpha^2 + \beta^2)
\ , \\ \label{ed4a}
	\frac{ 2 \, A \, r } { T } \, \frac{ dT } { dr } 
	&= & A - 1 - 8 \, \pi \, G \, T \, N \, \times
\\  
	& & 
	\left( \omega \, T  \, (\alpha^2 + \beta^2)  - \sigma \,N \, \frac{ \alpha \, \beta}{r} \right)
\ . \nonumber 
%	\\
\end{eqnarray}
These equations have a power-law solution of the form
\bea
\alpha = \alpha_p \, r, \quad
\beta = \beta_p\, r, \quad
A = A_p, \quad
T = \frac{T_p}{r}, \label{plsol}
\eea
with the coefficients $\alpha_p$, $\beta_p$, $A_p$, and $T_p$ given by
\bea
	\alpha_p^2 & = &	
	\frac{ \omega } { 12 \, \pi \, G \, \sigma \, N^2 \, N_- } 
\ ,  \label{plsolparam1} 
\hspace{5mm} 
	\left( \frac{ \beta_p } { \alpha_p } \right)^2
	= \frac{ N_- } { N_+ } 
\ , \\[12pt]
	A_p & = & \frac{ 1 } { 3 }
\ , 
\hspace{35mm} T_p =
	\frac{ \overline{ N } } { \omega }
\ , \label{plsolparam4}
\eea
where  we define $ \displaystyle N_\pm \equiv \frac{ \sigma\, N} { 2 } \pm \sqrt{ A_p }$ and
\be
\overline{N} \equiv \left( N_+ \, N_- \right)^{1/2}
	=  \left( \frac{ N^2}{4} - \frac{ 1 } { 3  } \right)^{1/2}
\ .
\ee
Note that while $T(r)$ is dimensionless, $T_p$ is the radius at which the power-law $T(r)=T_p/r=1$.
The fermion number density is
\be
	\eta = \frac{N \, \left( \alpha^2 + \beta^2 \right)\, T}{r^2}
	= \frac{1}{12\,\pi\, \overline{N} \, G\, r}
\ .
\ee

This power-law solution might be considered unphysical in two senses.  
First, the solution is singular at the origin,
where the gravitational redshift  $T(r)-1$ 
and the fermion number density $\eta(r)$ both diverge as $1/r$.
Second, the normalization integral for the fermion wave function,
\be
	4\,\pi \int\limits_0^\infty \frac{ \left( \alpha^2 + \beta^2 \right) \, T \, dr }{ \sqrt{ A } } 
%= \frac{ 2\, \pi \, \left( \alpha_0^2 + \beta_0^2 \right) \,  T_0 \, R^2 } { \sqrt{ A_0 } } 
= \lim\limits_{r\rightarrow\infty} \frac{ r^2 } { 32 \, G } \left( 1 - \frac{ 4 } { 3 \, N^2 } \right)^{1/2}
\ ,
 \label{normint}
\ee
diverges at large $r$, and hence the state is not normalizable. Despite these concerns, we shall show in the following sections that this pure power-law solution is nonetheless a good starting point for understanding the complete set of solutions of the FSY equations of motion (\ref{ed1}--\ref{ed4}).

\begin{figure*}[t]
\begin{center}
 \includegraphics[scale=1.]{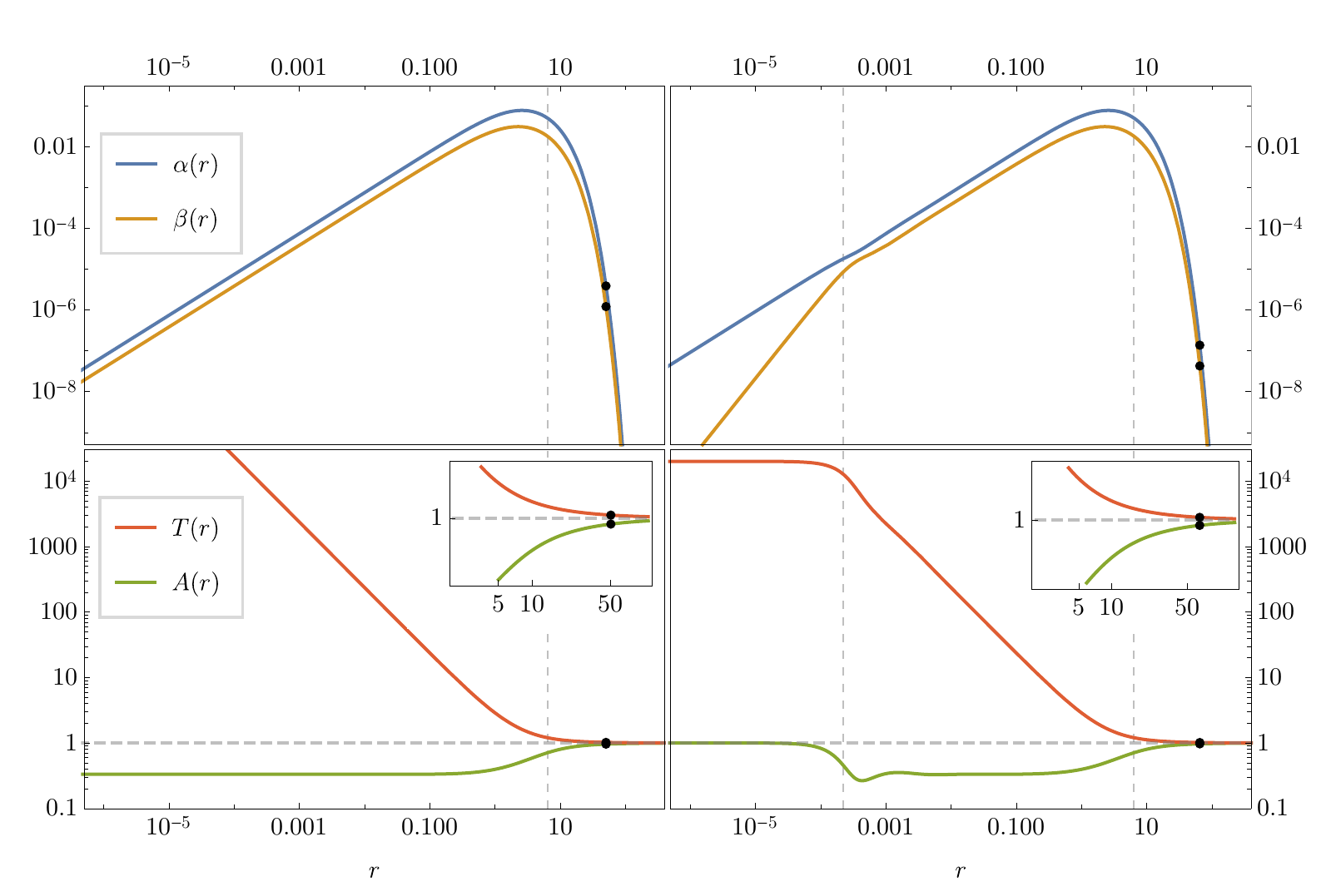} 
\end{center}
\caption{{\bf Left:} An infinite-redshift solution for $N=2$ fermions localized in an asymptotically flat spacetime.  
The parameters are 
$\left( z, m, \omega, M \right)=
	\left( \infty
	, 0.41508 %0.41508%16409688405640816516
	, 0.34522 %0.34521%811165042195070
	, 1.01079 %1.01079%0748284676278
	\right)$.
{\bf Right:} A finite-redshift ($z=20028$) solution for $N=2$ fermions
in an asymptotically flat spacetime, with parameters
$\left( z,m, \omega, M \right)=
	\left(20,027 %20027.46805983010472593208
	 ,0.41507 %0.41507%17832209226627984635
	,0.34521 %0.34521%2855828884857535506
	, 1.01076 %1.01076%130541010972858789
	 \right)$.
{\bf Upper panels:} Particle-like ($\alpha$) and hole-like ($\beta$) parts of the one-fermion wave function. 
 {\bf Lower panels:} Time-dilation ($T$) and length-contraction ($A$) metric parameters.
Black dots mark the points at which the numerical solutions are matched to
analytic large-$r$ asymptotic expansions.
Vertical dotted lines mark the transition radii
at $R\sim\omega\, T_p/m$ 
between the power-law and evanescent zones,
at the point where $\omega\,T(r)=m$,
and for the finite-redshift solution 
at $r_0=T_p/\left(1+z\right)\ll1$ 
between the core and power-law zones.
Note that $A=1$ and $T=1+z\approx2\times10^4$ in the core,
transitioning to $A=1/3$ and $T\approx 1/r$ in the power-law zone.
%{\bf Daniel to fill in the parameter values.}
}
\label{fig:1}
\end{figure*}

\begin{figure*}[t]
\begin{center}
 \includegraphics[scale=1.]{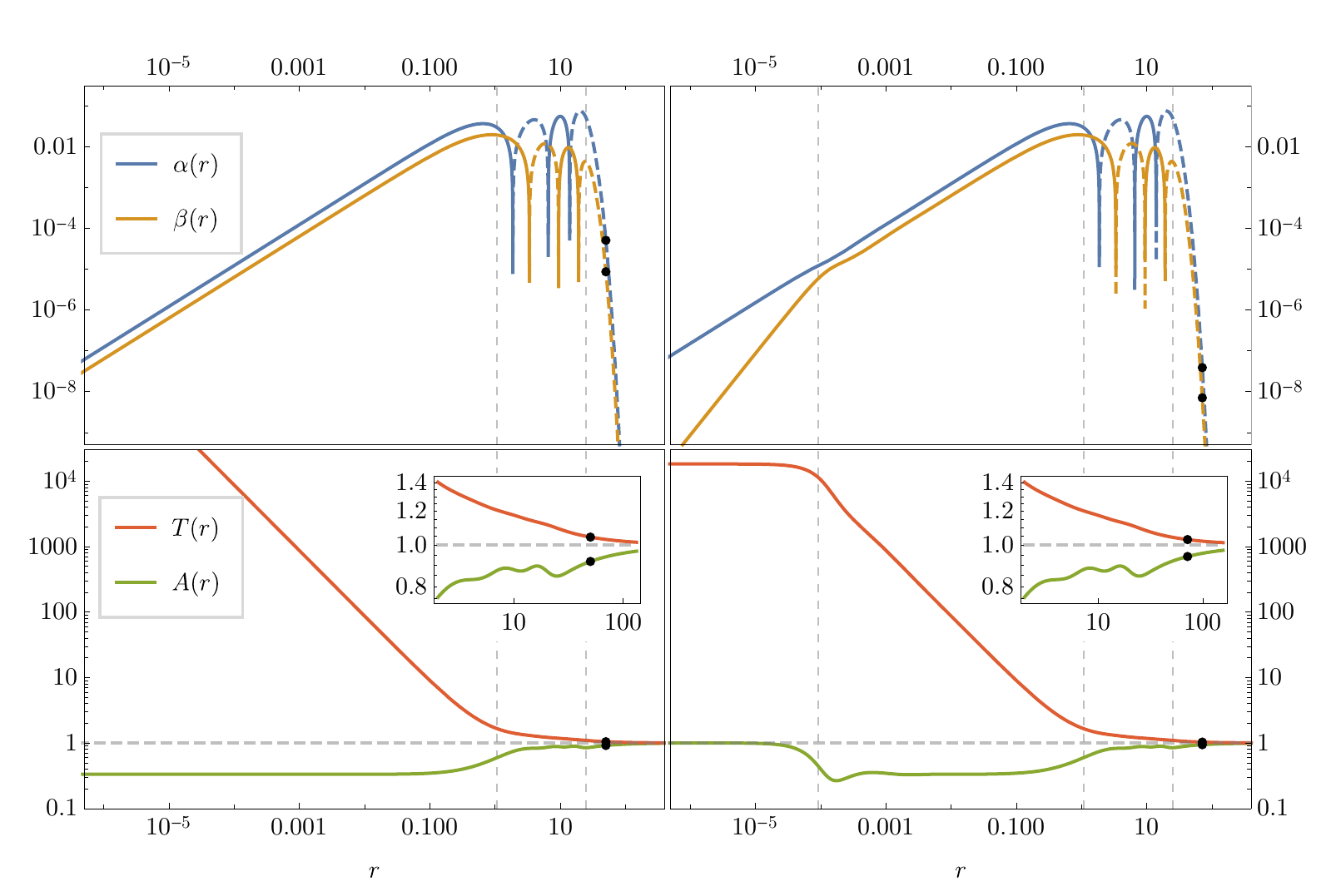} 
\end{center}
\caption{As in Fig.~\ref{fig:1}, solutions for $N=2$ fermions localized in an asymptotically-flat 
spacetime, 
but now for an excited state with $n=6$ fermion wavefunction nodes,
3 zeros of $\alpha(r)$ alternating with 3 of $\beta(r)$, forming a wave zone
between the power-law and evanescent zones.
{\bf Left:} An excited-state infinite-redshift solution, with 
parameters $\left( z,m, \omega, M \right)=
	\left( \infty
	, 1.02837 %1.02836%79993383949186013801
	, 0.93877 %0.93877%10246807333305
	, 2.09894 %2.09893%6074799726747
	\right)$.
{\bf  Right:} An excited-state finite-redshift solution 
 with parameters $\left( z,m, \omega, M \right)
 	=\left( 18246%.41071817220917888643
	 ,1.02824 %1.02824%06000136105337742160
	, 0.93866 %0.93865%9325152525692398177
	, 2.09914 %2.09913%707504870323450346
	 \right)$.
{\bf  Upper panels:} Particle-like ($\alpha$) and hole-like ($\beta$) parts of the one-fermion wave function (dashed lines indicate negative values of the functions) . 
{\bf  Lower panels:} Time-dilation ($T$) and length-contraction ($A$) metric parameters.
Black dots mark the points at which the numerical solutions are matched to
analytic large-$r$ asymptotic expansions.
Vertical dotted lines mark the transition radii between the
core, power-law, wave and evanescent zones.
%{\bf Daniel to fill in the parameter values.}
}
\label{fig:2}
\end{figure*}

\section{An infinite-redshift solution with an asymptotically flat spacetime}
\label{s:asympflat}

To examine the relationship of our singular infinite-redshift solution to the regular localised states obtained by FSY, we begin by restoring a non-zero fermion mass $m$, which enables us to enforce the normalization of the fermion wave function, i.e.\ that the integral (\ref{normint}) should be unity.  This yields a solution that initially has two zones:\ an inner, `power-law' zone ($r \lesssim R$), where 
$\omega\,T>m$ and the fermion and metric fields approach the power-law form (\ref{plsol}); 
and an outer, `evanescent' zone 
($r \gtrsim R$), where $\omega\,T<m$ causes the fermion fields to decay exponentially,
allowing the metric fields to approach the Schwarzschild form,
\be
A = \frac{1}{T^2} = 1 - \frac{2\,G\,M}{r}
\ ,
\ee
where $M$ is the ADM mass of the localised state. Likewise, the fermion fields approach the asymptotic form
\be
	\frac{\alpha}{\alpha_\infty} \rightarrow \left( \frac{r}{a} \right)^b \, e^{-r/a}
\ , \hspace{5mm}
	\frac{\beta}{\alpha_\infty} \rightarrow \sqrt{ \frac{ N_- } { N_+ } } \, 
	 \left( \frac{r}{a} \right)^b \, e^{-r/a}
\ , 
\ee
where the exponential scale length $a$ and power-law index $b$ are given by
\be
	\frac{1}{a^2} = m^2 - \omega^2
	\ ,
	\hspace{5mm} b = \frac{G\,M}{a} \, \left( \omega^2 \, a^2 - 1 \right)
\ .	
\ee
(Appendix~\ref{appendixB} gives the next-order term in the large-$r$ 
approximations to $\alpha(r)$ and $\beta(r)$.)
The transition between these zones occurs at a radius $R$, where $\omega\, T(R) = m$, which can be estimated
\be
	m \, R \approx \omega\, T_p = \overline{N}
\ .
\ee

Because we need to fine-tune the fermion frequency $\omega$ to avert a large-$r$ runaway of the 
fermion fields, and thus divergence of the normalization integral (\ref{normint}), such solutions must be obtained numerically. We use a shooting method to find $\omega$ and rescaling of the fields and parameters to impose the normalization, 
similar to the method outlined in FSY (see Appendix~\ref{appendixB} for details).

The left-hand panels of Fig.~\ref{fig:1} show the fermion and metric fields for such a solution, in this case the lowest-energy two-particle state.  The transition radius $R$ between the inner power-law zone and the outer evanescent zone
is marked by a vertical dotted line, and a black dot on each curve marks the radius at which
we match the numerical solution onto large-$r$ asymptotic expansions.

%{\bf CAH: We should add, here or in the caption (or both), the parameters needed to generate this solution.  If we have done some matching to asymptotic forms when generating the graph, that should be mentioned too.}

\section{Infinite-redshift solutions with multiple zeros in the fermionic wave function}
\label{s:wavezone}
\begin{figure}[]
\begin{center}
 \includegraphics[scale=1.]{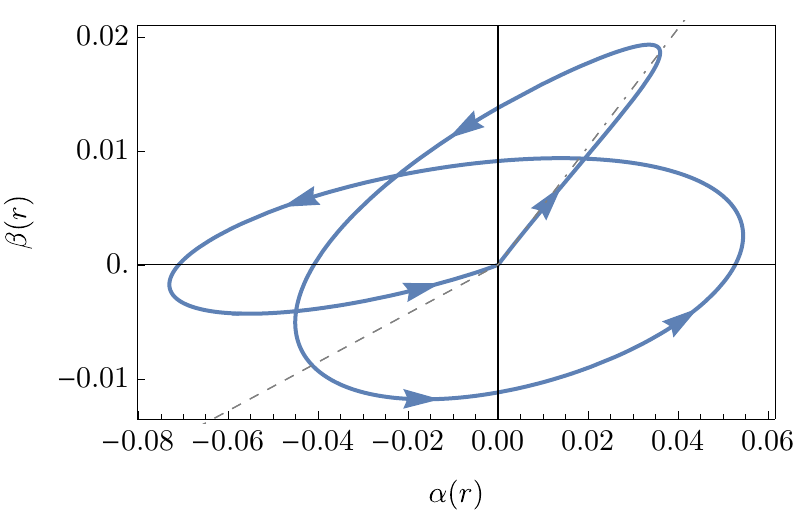} 
\end{center}

\caption{ Trajectory in the $(\alpha,\beta)$ plane
for the $N=2$ fermion excited-state solution
with central redshift $z\approx2\times10^4$ and $n=6$ axis crossings,
as shown in the right panel of Fig.~\ref{fig:2}. Arrows indicate increasing radius.
In the core ($r\lesssim10^{-4}$, $A\approx1$, $T\approx2\times10^4$)
the fermion field moves from the origin into the 1st quadrant with
$\alpha\propto r$ and $\beta \propto r^2$.
This rise continues in the power-law zone
($10^{-4}\lesssim r \lesssim1$, $A\approx1/3$, $T\approx1/r$)
where $\alpha$ and $\beta$ increase together, both $\propto r^2$,
with a ratio close to the dashed line 
with the slope $\beta/\alpha=\sqrt{ N_+ / N_+ } 
= \sqrt{ \sqrt{3} - 1 } / \sqrt{ \sqrt{3} + 1 }$.
With $T$ decreasing, the solution transitions to the
wave zone ($1\lesssim r \lesssim 20$, $T\sim1$, $A\approx0.9$), 
where $\omega\,T>m$ drives a counter-clockwise circulation
tracing a series of ellipses and crossing an axis 6 times 
to arrive in the 3rd quadrant.
Here we encounter the evanescent zone ($\omega\,T<m$) 
where $\alpha$ and $\beta$ decay exponentially with a ratio approaching the dotted line
with slope $\beta/\alpha=\sqrt{m-\omega}/\sqrt{ m+\omega }$.
%{\bf 
%Ideally 
%colour-code the trajectory to indicate transitions from core to power-law to wave to evanescent zones.}
\label{fig:alphabeta}
}
\end{figure}

As in the case of the finite-redshift solutions studied by FSY, the infinite-redshift solutions form an excitation spectrum of which the solution on the left of Fig.~\ref{fig:1} is the ground state.  While the ground-state solution has only two zones (power-law and evanescent), the associated excited-state solutions also exhibit a `wave zone' in between the power-law and evanescent zones.  In this wave zone the fermion wave function exhibits oscillations qualitatively similar to those that would be seen in the excited states of a non-relativistic quantum particle confined by a linear potential.  An example of such an excited-state solution is shown in the left-hand panels of Fig.~\ref{fig:2}, with the boundaries between the three zones marked by vertical grey dashed lines.  
In this example,  3 zeros of $\alpha(r)$ alternating with 3 of $\beta(r)$ occur in the wave zone
inserted between the inner power-law zone and the outer evanescent zone.
This corresponds to a counter-clockwise circulation in the $\left(\alpha,\beta\right)$ plane,
as seen in Fig.~\ref{fig:alphabeta}.

The transition from power-law to oscillatory behavior occurs at 
the transition from high to low redshift, roughly at  $r \sim T_p$.
We estimate the transition radius as
\be
	r_{w} 
\approx  \ds
	\frac{1}{m} 
	\left( \frac{ N } { 2 }
	\left( \frac{\alpha_p}{\beta_p}\right)^\sigma - \sigma \, \omega \, T_p \right)
%\\[4pt]
%= \frac{1}{ m} \: \left( \frac{ \sigma \, N } { 2 } \sqrt{\dfrac{N_+}{N_-}}- \overline{N} \right) 
= 
	\ds \frac{1}{m} \left( \frac{ N + 2/\sqrt{3} }{ 3 \, N - 2 \, \sqrt{ 3 } } \right)
\ . 
\ee
This is the first peak of $\alpha$ if $\sigma=+1$, or of $\beta$ if $\sigma=-1$, 
which marks the boundary between the power-law and wave zones in our figures.
In these excited states, the wave zone occurs because the increased $\omega$
relative to the ground state allows the highly-relativistic power-law
zone, with $T\gg1$, to become sub-relativistic, with $T\sim1$ and $\omega\,T>m$,
before entering the evanescent zone, where $\omega\,T<m$. 
The oscillatory behavior in the wave zone
can be deduced from the Dirac equations (\ref{ed1})-(\ref{ed2}) in the limit $r \gg1$:
\bea
\label{eqn:nonrel}
	\sqrt{A} \,\frac{d\alpha}{dr} &\approx & - (E+m) \, \beta\ , \\
\label{eqn:nonrel2}
	\sqrt{A} \,\frac{d\beta}{dr} &\approx & + (E-m) \, \alpha\ . 
\eea
Both metric fields $A(r)$ and $T(r)$ tend to unity for large $r$ and so, qualitatively at least, the above equations can be seen as an oscillator system so long as $E(r)\equiv \omega\,T(r)>m$. 
The radius outside which this happens defines the boundary between the oscillatory and evanescent behavior of the wavefunction.
In the wave zone, the fermion fields circulate in the $(\alpha,\beta)$ plane,
with a radial wave number
\begin{equation}
	k(r) = \left( \frac{ E(r)^2-m^2 } { A(r) } \right)^{1/2}
\ ,
\end{equation}
tracing an ellipse whose aspect ratio becomes more flattened as $r$ increases, as shown in 
Fig.~\ref{fig:alphabeta}.

For the ground state ($n=0$) and these excited states ($n>0$), the quantum number $n$
counts the number of times the fermion fields cross the $\alpha$ and $\beta$ axes 
in the wave zone before their exponential decay after 
crossing into the evanescent zone, where $E(r)<m$.
The example in Fig.~\ref{fig:2}, with 3 zeros of $\alpha$ and 3 of $\beta$,
is the excited state with $n=6$.
We note that $n$ is even for $\sigma=+1$ and odd for $\sigma=-1$.

The evanescent zone, where $E(r)<m$, requires $\alpha$ and $\beta$ to
have the same sign so that the $m$ term in the Dirac equation 
(\ref{eqn:nonrel})-(\ref{eqn:nonrel2})
drives exponential decay of the fermion fields.
 Thus the solutions must end in the 1st or 3rd quadrants of 
the $\left(\alpha,\beta\right)$ plane.
The ground state, $n=0$ begins and ends in the first quadrant,
and is dominated by $\alpha$ at small $r$.
The $n=1$ state begins in the 4th quadrant,
with $\beta<0$ dominating $\alpha>0$ at small $r$,
then crosses into the 1st quadrant before the evanescent decay.
The $n=2$ state starts in the 1st, circulates through the 2nd
and ends in the 3rd quadrant.
The $n=3$ state starts in the 4th, circulates through the 1st and 2nd,
and ends in the 3th quadrant.
In this way we have a countably-infinite tower of infinite-redshift excited states each
with a singluar power-law zone, a wave zone with $n$ fermion nodes,
and an evanescent decay.

%\textbf{DBC: How much detail should we give on the wave zone? Bzw evanescent zone.
%KDH: Illustrate with Fig.~\ref{fig:alphabeta} showing the ellipses in the $(\alpha,\beta)$ plane.}

\section{The relationship to FSY's finite-redshift solutions}
\label{s:fourzones}

We now relate our infinite-redshift solutions to the finite-redshift solutions obtained by FSY by restoring the boundary condition that all fields  in (\ref{ed1})-(\ref{ed4}) should be regular at $r=0$. Assuming that each field can be expanded in a Maclaurin series around $r=0$, this implies the following asymptotic behavior at small $r$:
\bea
	\alpha(r) & = &
	r^{N/2} \left( \alpha_0 + \alpha_1\, r + \mathcal{O}(r^2) \right)
\ ; \label{smallr1} \\
	\beta(r) & = & r^{N/2} \left( \beta_0 + \beta_1 \, r + \mathcal{O}(r^2) \right); \\
	A(r) & = &
		1 + \mathcal{O}(r^{N}); \\
	T(r) & = & 
	T_0 +\mathcal{O}(r^{N}). \label{smallr4}
\eea
(We note that $\alpha_0>0$, $\alpha_1=0$, $\beta_0=0$,
and
\be 
	\frac{ \beta_1}{\alpha_0}= \frac{\omega\, T_0 - m}{N+1}
\ee
for $\alpha$-dominated states with $\sigma=+1$
and $n$ even,
while $\beta_0<0$, $\beta_1=0$, $\alpha_0=0$, and
\be
	\frac{\alpha_1}{\beta_0} =  - \frac{\omega\, T_0 - m}{N+1}
\ee
for $\beta$-dominated states with $\sigma=-1$
and $n$ odd.)

These non-singular solutions have a finite central redshift, $z=T_0-1$.
\begin{figure}
\begin{center}
 \includegraphics[scale=1.]{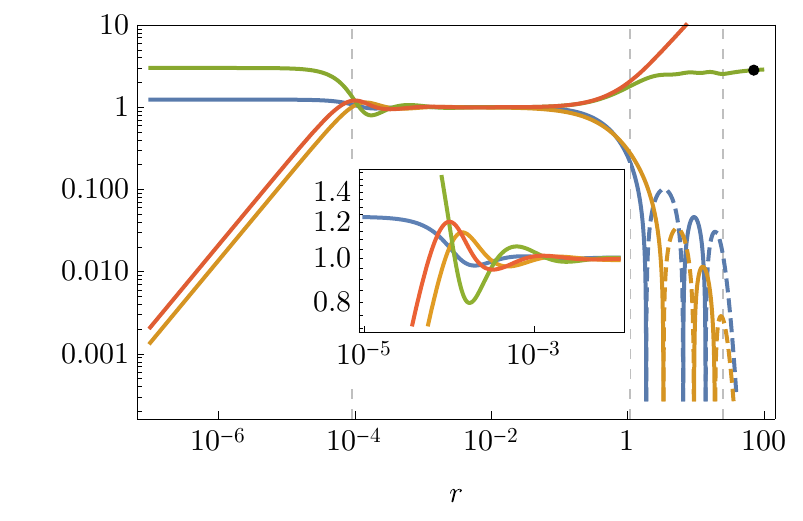} 
   \includegraphics[scale=1.]{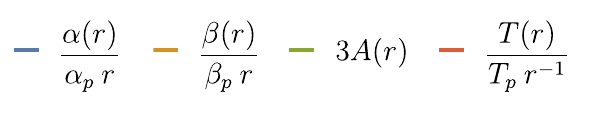} 
\end{center}
\caption{A numerical solution of the FSY equations with finite central redshift ($z\approx2\times10^4$) embedded in an asymptotically flat spacetime, plotted relative to the pure power-law solution.  Note the extended region of $r$ in which the numerical solution follows the pure power-law form.
The damped oscillations, highlighted in the inset, are discussed in Appendix~\ref{appendixC}.
}
\label{fig:4}
\end{figure}
Imposing the boundary conditions (\ref{smallr1})--(\ref{smallr4}) introduces a new zone in the centre of the soliton, which we call the `core zone'.

\begin{figure}[t]
\begin{center}
\hspace{.9cm}\includegraphics[scale=1.]{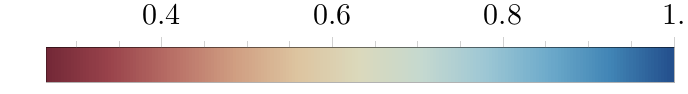}
\includegraphics[scale=1.]{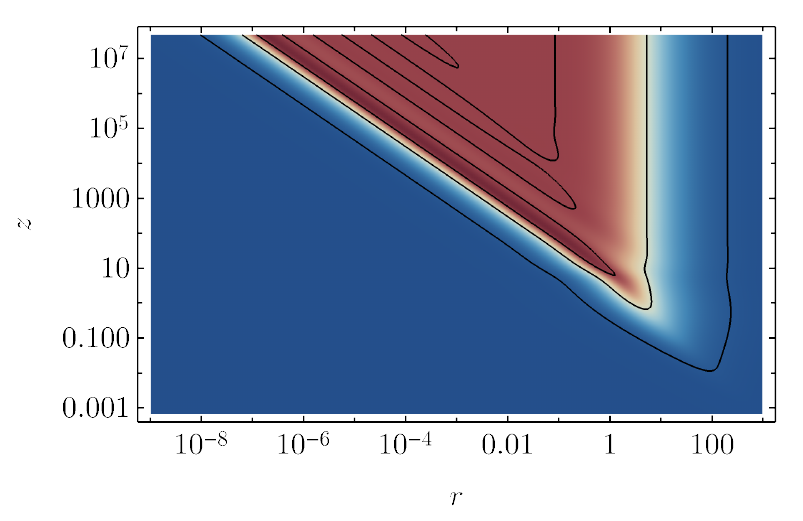}
\end{center}
\caption{Phase diagram for the ground-state $N=2$ solution, showing
the length-contraction metric parameter $A(r,z)$
as a function of radius $r$ and central redshift $z$.
This log-log plot covers a wide range of $r$ and $z$.
Contours are drawn at $A=(0.99,2/3,1/3)$.
Regions of approximately flat space ($A\approx1$, red) include the core at $r< r_0\sim1/(1+z)$
and the evanescent zone at $r>R\sim1$.
The power-law zone ($A\approx1/3$, red) occupies the
 triangular region between $r_0\sim1/z$ and $R\sim1$, with an apex near $z\sim1$.
The $A=1/3$ contour shows the damped oscillations of $A$ above and below its pure power-law value,
launched by the overshoot at the inner edge of the power-law zone.
Because the core radius $r_0\propto1/(1+z)$, the
power-law zone oscillations in $\ln{r}$ at fixed $z$ correspond to oscillations in $\ln{z}$ at fixed $r$.
\label{fig:5}
}
\end{figure}

\begin{figure}[t]
%\begin{flushleft} {\bf (a) } \end{flushleft}
\begin{center} 
{\bf(a)} \hspace{4mm}\includegraphics[scale=1.]{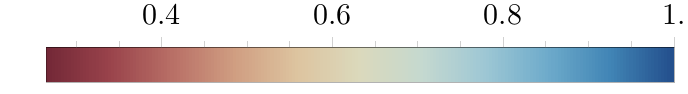}
\includegraphics[scale=1.]{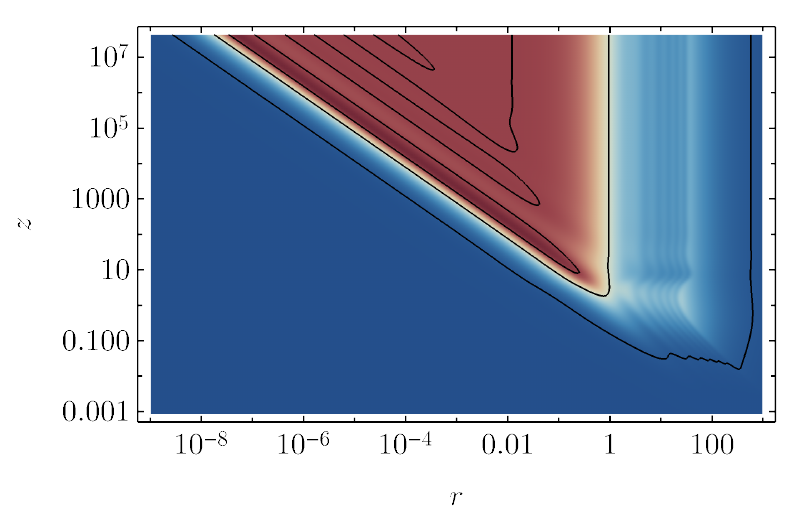}
{\bf(b)}
\hspace{4mm}\includegraphics[scale=1.]{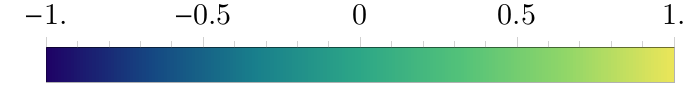}
\includegraphics[scale=1.]{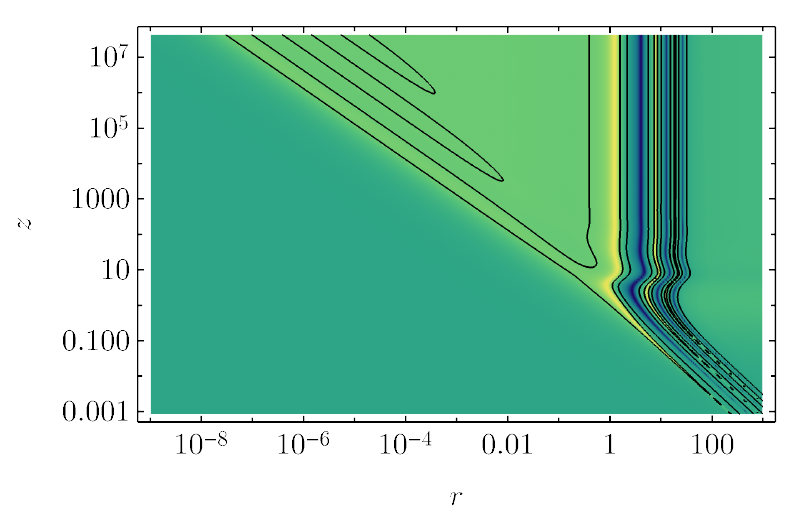}
\caption{ Phase diagram for the $N=2$ fermion excited-state solution
with $n=12$ nodes in the fermion wave function.
{\bf (a):} The length-contraction metric field  $A(r,z)$ 
as a function of radius $r$ and central redshift $z$, 
with contours at $0.99$, $2/3$ and $1/3$.
Comparison with Fig.~\ref{fig:4} shows that for $r<1$
the core ($A\approx1$,blue) and power-law ($A\approx1/3$, red) zones are
essentially unchanged. The excitation inserts
a wave zone ($A\approx0.9$, light blue) between the power-law and evanescent zones.
{\bf (b):}  $\beta / \sqrt{\alpha^2 + \beta^2}$ with contours at 0,
to highlight the wave-zone oscillations in $r$,
and at $\sqrt{ \frac{1}{2} - \frac{ 1 } { 2 \, \sqrt{3} } } \approx0.459$,
to highlight the power-law zone damped oscillations in $\ln{r}$.
\label{fig:6}}
\end{center}
\end{figure}

In the core zone, for the case where the redshift is reasonably large ($z \gtrsim 1$), the fields $\alpha$, $\beta$, $A$, and $T$ interpolate between their $r=0$ values and the functional forms given by the power-law solution (\ref{plsol}). The transition occurs at the radius 
\be
	r_0\approx T_p/T_0=T_p/(1+z).
\ee
or alternatively
\be
	r_0\approx \frac{N+1}{E_0-m} \sqrt{\frac{N_-}{N_+}},
\ee
obtained by comparing the ratio of the fermion fields  from both power law and core predictions. 
%{\bf DBC: $r_0\propto1/(1+z)$ is not a good match to solutions. Throw away or keep?}
This change in behavior is shown in the right-hand panels of Fig.~\ref{fig:1} (for the ground-state case) and Fig.~\ref{fig:2} (for the $n=6$ excited-state case).  Here, in addition to the boundaries between zones that were visible in the infinite-redshift case, there is now a new boundary  at $r_0 \approx 10^{-4}$ between the core and power-law zones.

Fig.~\ref{fig:4} shows a comparison between the excited-state finite-redshift solution in the right-hand panels of Fig.~\ref{fig:2} and the pure power-law solution (\ref{plsol}), created by plotting the fermion and metric fields divided by their power-law values. We see that, over many decades in $r$, the two solutions match essentially perfectly.  Note that, at the boundary between the core and power-law zones ($r_0\approx 10^{-4}$), the core-zone solution `overshoots' the infinite redshift power-law solution, creating oscillations in the fermion and metric fields that die away with increasing radius.  These can be described quantitatively as perturbative corrections to the power-law solution, a topic which is further developed in Section~\ref{s:poweroscillations}
and Appendix~\ref{appendixC}.

The size of the core zone, $r_0$,  is determined by the distance needed for the fields at $r<r_0$ to reach their power-law values, which is in turn determined by the redshift $z$.  As $z$ decreases, the core zone expands,
replacing the inner part of the power-law zone. At $z\sim1$ the power-law zone disappears and the core zone subsequently connects directly to the wave zone (or evanescent zone in the ground-state case) without any intervening power-law behavior.  This is illustrated in the phase diagram, Fig.~\ref{fig:5}, which has been constructed by combining the graphs of the length-contraction parameter $A(r)$ from ground-state $N=2$ solutions with a wide range of redshifts.    As the redshift is reduced, the core--power-law boundary at $r_0\approx1/(1+z)$
and the power-law--wave boundary at $R\approx1$  approach each other.
For redshifts $z \lesssim 1$ the power-law zone is eliminated entirely.

A phase diagram showing $A(r,z)$ for the $n=6$ excited-state solution is
presented in Fig.\ref{fig:6}.
Comparing the phase diagram showing $A(r,z)$ 
for the $N=2$ ground state in Fig.~\ref{fig:4} 
and that for the $n=6$ excited state in Fig~\ref{fig:5}a
indicates that the core and power-law zones are essentially unchanged,
but a wave zone with $A \approx 0.9$ forms outside $r\sim1$
between the power-law and evanescent zones for $z>1$
or between the core and evanescent zones for $z<1$.
The oscillating fermion fields are more clearly seen in
Fig.~\ref{fig:6}~b showing $\beta/\sqrt{ \alpha^2+\beta^2}$.
This indicates where the hole-like ($\beta$) component is significant.
$\beta$ is small in the core and evanescent zones.
In the power-law zone, the analytic estimate is
\be
	\frac{\beta^2}{\alpha^2+\beta^2}
	= \frac{ N_- } { N_+ + N_- } = \frac{ 1 } { 2 } - \frac{ \sigma } { \sqrt{ 3 } \, N } 
\ .
\ee
The damped oscillations in $\ln{r}$ are evident in the power-law zone.
In the wave zone $\beta$ oscillates in $r$ above and below 0,
with 3 maxima and 3 minima before the exponential decay in the evanescent zone.

\section{Oscillations around the power law solution: origin of the FSY spirals}
\label{s:poweroscillations}

\begin{figure}
\begin{center}
\includegraphics[scale=1.]{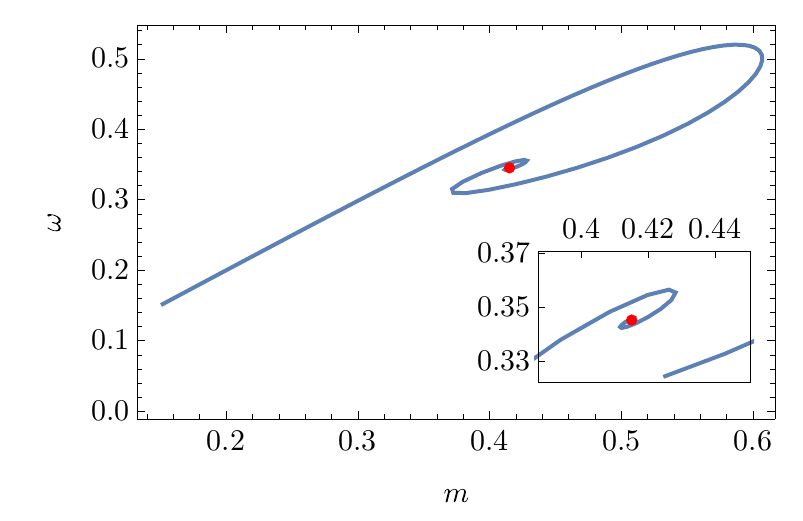}
\end{center}
\caption{The fermion frequency, $\omega$, as a function of fermion mass, $m$, for 
the family of ground-state solutions to the FSY equations.  
As the central redshift is increased, points ever closer to the center of the spiral are generated.  
Thus the very center corresponds to the infinite redshift case, i.e.\ 
the one in which the power-law zone expands all the way to $r=0$. The red dot marks the $(m,\omega)$ value of an infinite red-shift soliton for the ground state. The predicted co-ordinates $(0.4151, 0.3452)$ match the centre exceedingly well. }
\label{fig:7}
\end{figure}

\begin{figure}[t]
\begin{center}
\includegraphics[scale=1.]{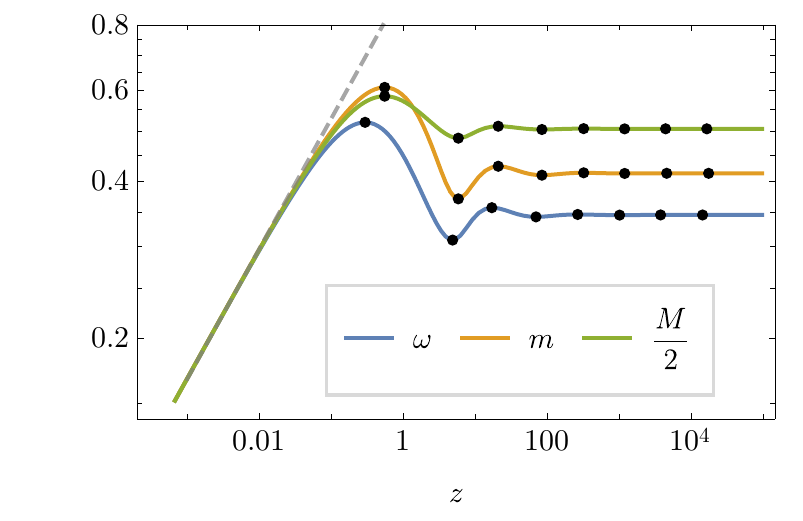} 
\end{center}
\caption{Dependence of the ground-state $N=2$ soliton parameters,
$m$, $\omega$, $M$, with central redshift $z$. Black dots mark extrema.
In the low-redshift limit, $z<<1$,  the fermion mass $m$, energy $\omega$ and the ADM mass 
per fermion $M/N$  all converge to the same power-law  $\omega\propto z^{1/4}$.
With redshift increasing toward $z\sim1$, all 3 parameters drop below the $z^{1/4}$ asymptote,
attaining maxima at slightly different redshifts with $\omega<M/N<m$.
The solitons are energetically bound up to the redshift where $M/N$ exceeds $m$.
and unstable to fragmentation at higher redshifts.
All 3 parameters undergo damped oscillations in $\ln{z}$ 
around different asymptotic values as $z\rightarrow\infty$,
where $\omega<m<M/N$.
The $\omega$ oscillations in $\ln{z}$ have a phase shift relative to those
of $m$ and $M$, resulting in the spiral pattern in the $m$ vs $\omega$ plane of Fig.~\ref{fig:7}.
}
\label{fig:8}
\end{figure}

As shown in Fig.~\ref{fig:7}, the high-redshift soliton properties, such as the fermion mass $m$ and the fermion frequency 
$\omega$, spiral around and converge upon the infinite-redshift limit (marked with a red dot in Fig.\ref{fig:7}).  This spiral arises because these soliton properties oscillate with increasing redshift before converging to their infinite-redshift values, as shown in Fig.~\ref{fig:8}; since the oscillations are not in phase with each other the result is a spiral rather than a straight line.  We note that, as shown in Fig.~\ref{fig:8}, these oscillations are periodic in $\ln z$ rather than in central redshift $z$.

These oscillations can be understood as a consequence of damped oscillations around the power-law solution that are
excited at the redshift-dependent core radius $r_0=T_p/(1+z)$.  It is clearly not possible, in the generic case, for the core-zone solution (\ref{smallr1}--\ref{smallr4}) to match on perfectly to the power-law solution (\ref{plsol}) at the radius $r_0$ where they meet.  There must therefore be a range of radii around that point where the solutions `adjust' from the core-zone to the power-law behavior.  We show in Fig.~\ref{fig:4} an example of this:\ the inset shows in more detail the small oscillations that the fermion and metric fields undergo relative to the pure power-law behavior (\ref{plsol}) as they enter the power-law zone.  These oscillations are periodic in $\ln r$, not in $r$, mirroring the logarithmic property of the oscillations in Fig.~\ref{fig:8}.

This behavior can be described in terms of small multiplicative corrections to the power-law solution:
\begin{eqnarray}
	\alpha(r) & = & 
	\alpha_p\, r \,\left( 1 + \epsilon\,{\overline{\alpha}}_1(r) + {\cal O}(\epsilon^2) \right)
\ , \\
	\beta(r) & = & 
	\beta_p \, r \,\left( 1 + \epsilon\,{\overline{\beta}}_1(r) + {\cal O}(\epsilon^2) \right)
\ , \\
	A(r) & = &
	A_p \, \left( 1 + \epsilon\,{\overline{A}}_1(r) + {\cal O}(\epsilon^2) \right)
\ , \\	T(r) & = &
	T_p \, r^{-1} \, \left( 1 + \epsilon\,{\overline{T}}_1(r) + {\cal O}(\epsilon^2) \right),
\end{eqnarray}
where $\alpha_p$, $\beta_p$, $A_p$, and $T_p$ are given in (\ref{plsolparam1}--\ref{plsolparam4}). 
Substituting these into the $m=0$ version of the FSY equations, (\ref{ed1a}--\ref{ed4a}), and neglecting terms 
${\cal O}(\epsilon^2)$, we obtain coupled differential equations for 
the corrections 
${\overline{\alpha}}_1(r)$, ${\overline{\beta}}_1(r)$, ${\overline{A}}_1(r)$, and ${\overline{T}}_1(r)$.

These equations are presented and solved in Appendix~\ref{appendixB}.  Of the four eigenmodes we obtain, one grows as a power law in $r$, one decays as a power law in $r$, and the remaining two show oscillations periodic in $\ln r$ as they decay.  This matches the periodicity in $\ln r$ of the oscillations seen in numerical solutions such as Fig.~\ref{fig:4}.

\begin{figure}[t]
\begin{flushleft}
{\bf (a)}
\end{flushleft}
\begin{center}
\includegraphics[scale=1.]{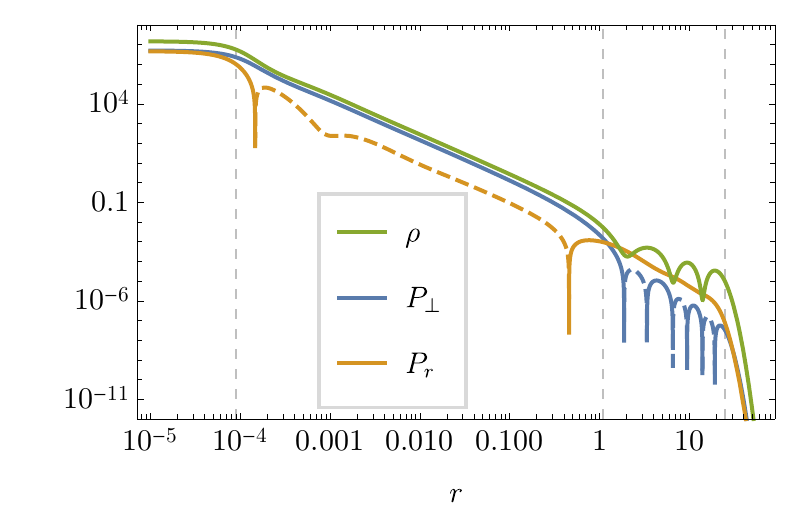} 
\end{center}

\begin{flushleft}
{\bf (b)}
\end{flushleft}
\begin{center}
\includegraphics[scale=1.]{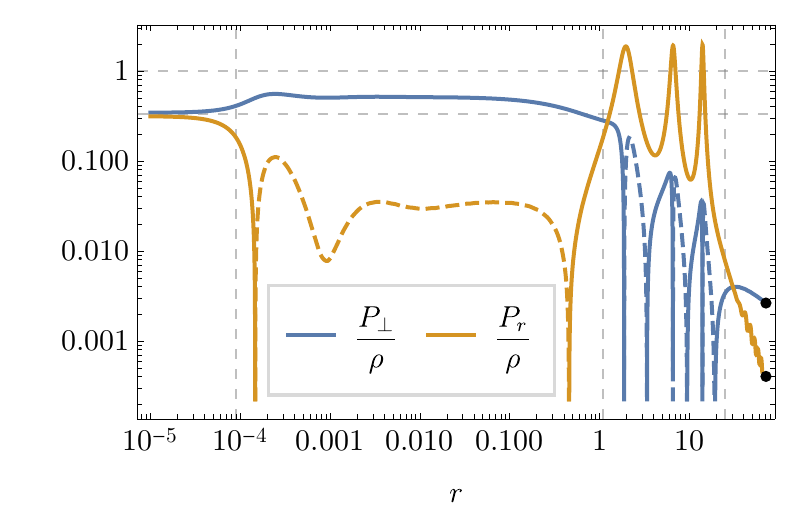} 
\end{center}
\caption{ {\bf (a)} Radial profile of the fermion energy density $\rho$,
radial pressure $P_r$, and azimuthal pressure $P_\perp$.
 {\bf (b)} Equation-of-state $P_r/\rho$ for the radial and 
 $P_\perp/\rho$ for the azimuthal pressures.
The parameters are $\sigma\,N=2$, $n=6$, and $z\approx2\times10^4$,
as in Figures \ref{fig:2} to \ref{fig:4}.
 Vertical dashed lines mark transitions from the
 core to power-law to wave to evanescent zones.
}
\label{fig:9}
\end{figure}

The damped oscillations in $\ln{r}$ induce damped oscillations in the 
high-redshift soliton properties $m$\, $\omega$, $M$, and $R$ as functions of redshift,
as shown in Fig.~\ref{fig:8}.
These arise as follows.
The oscillations in $\ln{r}$ are excited  at radius $r_0\approx T_p/(1+z)$, 
where the core-zone fields encounter and overshoot the power-law solution.
The perturbation analysis in Appendix~\ref{appendixB} indicates that the oscillation amplitude
then decreases as $1/r$, and thus the initial amplitude decreases by a factor $r_0/R$
at radius $R$ outside which $\omega\,T<m$, marking entry into the
evanescent zone.  Because the oscillating fields are perturbed above or below the 
power-law solution, so will be the radius $R$ at which they enter the evanescent zone,
thus altering the soliton properties $m$, $\omega$ and $M$.

\section{Physical properties of the fermion cloud}
\label{s:physprop}

The physical properties of the fermions in the FSY solitons provide another distinction between the four zones described above.  In Fig.~\ref{fig:9}, we plot the energy density $\rho$ and the radial and azimuthal pressures, $P_r$ and $P_\perp$ respectively, as functions of radius for 
the $N=2$ fermion $n=6$ excited-state solution with central redshift $z\approx2\times10^4$. 
Equation-of-state parameters, simply dividing the pressures by the energy density,
are shown in Fig.~\ref{fig:9}b.
The fermion stress-energy tensor, for the spherical-polar coordinates $x^\mu=(t,r,\theta,\phi)$, is
$T^\mu_\nu = {\rm diag}\left(\rho, -P_r\, -P_\perp, -P_\perp\right)$.
The fermion energy density is
\be
	\rho = \frac{ N \, \omega \, T^2 \,
	\left( \alpha^2 + \beta^2 \right)}{r^2} = \eta \, E
\ .
\ee
where $\eta$ is the fermion number density and $E \equiv \omega\, T$.
The radial pressure is
\be
	P_r
	= \frac{ N \, T } { r^2 } \,
	\left( \alpha^2 \, E_+ + \beta^2 \, E_- - \frac{ \sigma\, N \, \alpha\, \beta } { r } \right)
%	= \frac{N\,T}{r^2} \left( \omega\, T\, \left( \alpha^2 + \beta^2 \right)
%	- \frac{ \sigma\, N \, \alpha\, \beta} { r }
%	- m \left( \alpha^2 - \beta^2 \right) \right)
\ ,
\ee
where $E_\pm\equiv E\pm m$, and the azimuthal pressure is
\be
	P_\perp
	= \frac{ \sigma \, N^2 \,T\, \alpha\, \beta} { 2 \, r^3 }
\ .
\ee
The radial and azimuthal pressures behave quite differently in the four distinct zones.

The high-redshift core zone, $r \lesssim r_0\approx 10^{-4}$, 
is approximately uniform and isotropic.
The fermion energy density and pressure are independent of radius, 
and the fermionic matter is isotropic, i.e.\ the radial and azimuthal pressures are equal,
with $P/\rho=1/3$. Here the space is flat ($A\approx1$) and the fermions resemble an isotropic distribution of massless particles
moving at the speed of light.

 In the power-law zone, $10^{-4} \lesssim r \lesssim 1$, we have the
 fermion number density $\eta\approx1/\left(12\,\pi\,\overline{N}\,G\,r\right)$,
the energy per fermion $\omega\,T\approx \overline{N}/r$,
and thus the  energy density $\rho=\omega\,T\, \eta \approx1/ \left( 12\,\pi\, G \, r^2 \right)$. 
 The azimuthal pressure is $P_\perp\approx\rho/2$, and this
 dominates over $P_r$ by around two orders of magnitude.
The radial pressure actually vanishes in the pure power-law solution,
but here it oscillates above and below 0 as a result of the power-law zone
oscillations in $\ln{r}$.
Interestingly, for a metric with $T\propto1/r$
the classical circular orbit speed, given by $v^2=-d\ln{T}/d\ln{r}$, is the speed of light at all radii.
Here in the power-law zone the fermions resemble a collection of massless particles on circular orbits with an isotropic distribution of orientations.

%{\bf CAH: Presumably we can give exact expressions for the fermion density and the radial and azimuthal pressures from the power-law solution?  If so, perhaps we should.}  
%This represents a sort of `self-trapping' of the fermions on local extrema of the optical geometry --- see \cite{XXX} for more details. 

 In the wave zone, $1 \lesssim r \lesssim 20$, the energy density $\rho$ continues to decline but also starts to oscillate. Minima in $\rho$ correspond to sign changes in $P_\perp$ such
that $\rho$ increases with $r$ where $P_\perp<0$.
 The radial pressure $P_r$ is now dominant and declines monotonically,
 while the azimuthal pressure $P_\perp$ is small and oscillating around 0.
 Here the fermions resemble standing waves in a spherical cavity, bouncing 
between the interior power-law zone and the exterior self-generated gravitational potential.
 
 Finally, in the evanescent zone, $r \gtrsim 20$, the fermion energy density $\rho$ decays exponentially to zero and the pressures decrease even faster, becoming
 increasingly azimuthal, with $P_\perp/\rho\propto r^{-1}$  and $P_r/\rho\propto r^{-2}$.
Here the fermions resemble a collection of low-velocity test particles on elliptical orbits
with isotropic orientations and a range of radial turning points
such that fewer reach larger radii.

\section{Summary and outlook}
\label{s:summary}

In this paper, we have presented a power-law solution of the FSY equations for a filled shell holding
an even number $N$ of
neutral fermions interacting via the deformable spacetime metric of Einsteinian gravity.  This power-law solution does not obey the boundary conditions that FSY imposed:\ 
its central redshift is infinite, and its fermion wave function is not normalizable.

Nonetheless, it is in some sense a `key' to understanding the 4-zone structure of the full set of FSY solutions.  The evanescent zone at large $r$ (together with its preceding wave zone, where applicable) can be thought of as a deviation from the power-law solution due to the non-negligibility of the fermion mass $m$, which restores the normalizability of the fermion wave function.  Likewise, the core zone at small $r$ can be thought of as a deviation from the same power-law solution due to the requirement of a finite central redshift.

If we relax the requirement for a finite central redshift but keep the requirement that the fermion wave function be normalized, we obtain a set of infinite-central-redshift solutions that match the high-central-redshift limit of the known FSY solutions, i.e.\ that lie at the center of spirals like the one in Fig.~\ref{fig:7}.

We can go further:  The existence of such spirals is {\it explained by\/} the existence of the power-law solution.  As shown in Section~\ref{s:poweroscillations}, the failure of the core-zone solution to connect precisely to the power-law-zone one results in oscillations in the fermion and metric fields that are periodic in $\ln r$.  As we can see from the phase plots in Figs.~\ref{fig:5} and \ref{fig:6}, these oscillations follow the core--power-law boundary down to low red-shift, where they become oscillations of physical observables (such as the 
ADM mass and size of soliton) that are periodic in $\ln z$, where $z$ is the central redshift.

It would be interesting to extend this analysis to the case where more than one shell of fermions is filled.  This would provide a theory of gravitationally localised states that contain an arbitrarily large filled sphere of fermionic matter.  We expect the physics of such systems to have connections to cases previously studied in astrophysical contexts.  In particular, they should exhibit the same kind of gravothermal catastrophe discussed by Lynden-Bell and Wood \cite{LBW}, where spirals similar to those in Fig.~\ref{fig:7} and oscillations similar to those in Fig.~\ref{fig:8} are observed.  We plan to discuss this matter further in a future work.

\bibliography{FSY_Bib}{}
%\bibliographystyle{abbrv}

%
%
%\bibitem{FSY99}
%F. Finster, J. Smoller, and S.-T. Yau,
%{\it Phys. Rev. D\/} {\bf 59}, 104020 (1999).
%\bibitem{Weinberg}
%S. Weinberg, {\it Gravitation and cosmology: principles and applications of the
%general theory of relativity} (Wiley, 1972).
%
%\end{thebibliography}

\appendix

\section{Explicit expression for 2-spinors}
\label{appendixA}
The full 2-spinors are given by
\bea
	\chi^{k}_{j - \frac{1}{2}}(\theta,\phi) & = & 
	\sqrt{\dfrac{4 \pi (j+k)}{2j}} \, Y^{k-\frac{1}{2}}_{j-\frac{1}{2}}  (\theta,\phi)
	\left( \begin{matrix} \: 1\: \\ \: 0 \: \end{matrix} \right) \nonumber \\
&  + & \,\sqrt{\dfrac{4  \pi  (j-k)}{2j}} \, Y^{k+\frac{1}{2}}_{j-\frac{1}{2}} (\theta,\phi)
\left( \begin{matrix} \:0\: \\ \:1\: \end{matrix} \right)
\ , \\
	\chi^{k}_{j + \frac{1}{2}} (\theta,\phi) & = & 
	\sqrt{\dfrac{4 \pi (j+1-k)}{2j+2}} \, Y^{k-\frac{1}{2}}_{j+\frac{1}{2}} (\theta,\phi)
	\left( \begin{matrix} \:1\: \\ \:0\: \end{matrix} \right) \nonumber \\
& \bs\bs\bs\bs\bs\bs
	- & \bs\bs\bs
	\,\sqrt{\dfrac{4 \pi (j+1+k)}{2j+2}}  Y^{k+\frac{1}{2}}_{j+\frac{1}{2}} (\theta,\phi)
	\left( \begin{matrix} \:0\: \\ \:1\: \end{matrix} \right)
	\ ,
\eea
where $Y^m_l(\theta,\phi)$ denote the standard spherical harmonic functions. Differing from Finster et~al., an inconsequential factor of $\sqrt{4 \,\pi}$ has been introduced to the 2-spinors
so that the general system of equations reduces the original $N=2$ system given by \cite{FSY1}.

\section{Numerical solution of the FSY equations}
\label{appendixB}

For finite-redshift states, following FSY, we seek numerical solutions of the equations (\ref{ed1}--\ref{ed4}) in which the fermion fields $\alpha$ and $\beta$ tend to zero at large $r$ and the metric field $T$ tends to a constant.  We initialize the numerical solver at a small but non-zero radius $r = r_{\rm min}$ using the small-$r$ analytic forms of the unscaled solution.  These are similar to (\ref{smallr1}--\ref{smallr4}), but with unscaled parameters:
\bea
\alpha_{\rm SR}(r) & = & \alpha_{u0}\, r^{N/2}; \label{smallru1} \\
\beta_{\rm SR}(r) & = & \beta_{u1} \, r^{(N/2)+1}; \\
A_{\rm SR}(r) & = & 1; \\
T_{\rm SR}(r) & = & 1, \label{smallru4}
\eea
where
\be
\beta_{u1} = \frac{\omega_u - m_u}{N+1}\, \alpha_{u0}.
\ee
(The subscript `$u$' here indicates `unscaled' quantities.)
The free parameters in this small-$r$ solution are:\ $\alpha_{u0}$, which we choose arbitrarily, and which determines the central redshift of the eventual scaled solution; $m_u$, the fermion mass, which we set to $1$; and $\omega_{u}>1$, the fermion frequency.

To find a numerical solution in which the fermion fields tend to zero at large $r$, we tune the fermion frequency $\omega_u$ while keeping $\alpha_{u0}$ fixed.  Because this tuning can be carried out only to finite precision, our numerical solutions terminate at a finite radius $r_{\rm max}$, at which there is either an additional zero of one of the fermion fields (if $\omega_u$ is slightly too high) or an extremum preceding a divergence of the field to infinity (if $\omega_u$ is slightly too low).

Following FSY, we now aim to rescale our numerical solution to make the fermion fields normalized and to ensure that $\lim_{r \to \infty} T(r) = 1$.  However, this requires knowledge of the functions over the whole line $r \in \left( 0,\infty \right)$, whereas our numerical solutions cover only the finite interval $r \in \left( r_{\rm min},r_{\rm max} \right)$.  We therefore begin by matching our numerical solutions to the known small-$r$ and large-$r$ asymptotic forms. 

 The large-$r$ asymptotic expressions, 
in terms of a dimensionless radius $x\equiv r/a$, and the corresponding
dimensionless gravitational radius $\gamma \equiv G\,M/a$, are
\bea
\alpha_{\rm LR} (r) & = & \alpha_\infty \, x^b \,e^{-x} \,
	\left( 1 + \frac{\alpha_{e1}}{x} \right); \\
\beta_{\rm LR} (r) & = & \alpha_\infty \, x^b \, e^{-x} \, 
	\sqrt{\frac{m_u-\omega_u\,\tau}{m_u+\omega_u\,\tau}}
	 \, \left( 1 + \frac{\beta_{e1}}{x} \right); 
\,\,\,\,\,\,\,\,\,\,\,\, \\
A_{\rm LR} (r) & = & 1 - \frac{ 2 \, \gamma } { x}\ ; \\
T_{\rm LR} (r) & = & \tau \left( 1 - \frac{ 2 \, \gamma } { x } \right)^{-1/2}\ .
\eea
Here the fermion decay scale $a$ and leading exponent $b$ are
\bea
   a & = & \left( m_u^2 - \omega_u^2 \, \tau^2 \right)^{-1/2}\ ; 
\\ b & = & \gamma\, \left( \omega_u^2 \, \tau^2 \, a^2 - 1 \right)
\ .
\eea
The sub-leading coefficients of the expansion in powers of $1/x$ are
\be
\alpha_{e1}  =  \frac{\Sigma_{e1} + \Delta_{e1} }{2} 
\ , \hspace{5mm}
  \beta_{e1}  =  \frac{ \Sigma_{e1} - \Delta_{e1}}{2}  
\ ,
\ee
where $\Sigma_{e1}$ and $\Delta_{e1}$ are given by
\bea
   \Delta_{e1} & = & \gamma \, a^2 \, m_u \, \omega_u \, \tau - \frac{ \sigma \, N } { 2 }
\\ \Sigma_{e1} & = & \frac{N^2}{4} + \gamma^2 
	\left( 3 - a^2 \, \omega_u \, \tau \, \left( 5 \, \omega_u \, \tau + m_u \right) \right)
\,\,\,\,\,\,\,\,
\eea
(The subscript `$e$' here stands for `evanescent'.)

Matching the small-$r$ expressions to our numerical solution is unproblematic, since we used them as the boundary conditions for our numerical solution at $r=r_{\rm min}$.  However, to match our numerical solution to the above large-$r$ expressions we must determine the three unknown parameters $M$, $\tau$, and $\alpha_\infty$.

Since the errors in the fields $T$ and $A$ at $r_{\rm max}$ are much smaller than those in the fields $\alpha$ and $\beta$, we first determine $M$ and $\tau$ by matching the numerically determined $A$ and $T$ fields to their large-$r$ expressions at $r_{\rm max}$.  Specifically, we estimate $\tau$ and $M$ as
\bea \label{eqn:tau}
\tau & = & \left. \sqrt{A_{\rm num}(r)}\,T_{\rm num}(r) \right\vert_{r=r_{\rm max}}; \\
M & = & \left. \frac{r}{2} \left( 1 - A_{\rm num}(r) \right) \right\vert_{r=r_{\rm max}},
\eea
where $A_{\rm num}$ and $T_{\rm num}$ are the solutions for $A$ and $T$ given by the numerical solver.

Matching the fermion fields at $r_{\rm max}$ would be unsafe, since there they are strongly affected by our small error in the value of $\omega_u$.  Therefore, we instead look for a radius inside $r_{\rm max}$ where the numerically determined $\alpha$ and $\beta$ approach most closely their analytically determined large-$r$ ratio, i.e.\ where
\be
\left( \beta_{\rm num}(r) - \sqrt{\frac{m_u - \omega_u \tau}{m_u + \omega_u \tau}} \left( \frac{x + \beta_{e1}}{x + \alpha_{e1}} \right) \alpha_{\rm num}(r) \right)^2 \label{matchmin}
\ee
is minimized.  Calling this radius ${\tilde r}_0$, we actually do the matching of the fermion fields at a somewhat lower radius, $r_0 \equiv 4\, {\tilde r}_0 / 5$.  This is because (\ref{matchmin}) may actually become zero if the numerically determined ratio crosses the analytically determined one:\ that is clearly a minimum, but it occurs at a radius where the numerically determined fermion fields have already begun to diverge from the correct large-$r$ asymptotic form.  Having thus determined $ \tilde{r}$, we set $\alpha_{\rm num}(r_0) = \alpha_{\rm LR}(r_0)$, thereby determining $\alpha_\infty$.  This means that there is a slight jump in $\beta(r)$ at $r_0$, but this is an unavoidable consequence of the slight residual error in our numerical solutions.

These matching procedures yield a set of four functions, $\left\{ \alpha_{u}(r), \beta_{u}(r), A_u(r), T_u(r) \right\}$, which are defined on the whole interval $r \in \left( 0,\infty \right)$.  These do not yet satisfy conditions (i) and (ii):\ although the fermion fields are normalizable, they are not yet normalized, and although the field $T_u(r)$ has a finite value as $r \to \infty$, that value is not 1.  To remedy this, we follow FSY and define scaled versions of the fields:
\bea
\alpha(r) & \equiv & \sqrt{\frac{\tau}{\lambda}}\,\alpha_{u} \left( \lambda r \right); \label{scaled1} \\
\beta(r) & \equiv & \sqrt{\frac{\tau}{\lambda}}\,\beta_{u} \left( \lambda r \right); \\
A(r) & \equiv & A_u \left( \lambda r \right); \\
T(r) & \equiv & \frac{1}{\tau} \,T_u \left( \lambda r \right), \label{scaled4}
\eea
where
\bea
\lambda & = & \left( 4 \pi \int\limits_0^\infty \left( \alpha_u^2 + \beta_u^2 \right) \frac{T_u}{\sqrt{A_u}} \, dr \right)^{1/2},
\eea
and $\tau = \lim_{r \to \infty} T_u(r)$ was obtained  in (\ref{eqn:tau}) 
during the matching procedure.  The functions (\ref{scaled1}--\ref{scaled4}) satisfy the equations (\ref{ed1}--\ref{ed4}) but for different values of the fermion mass $m$ and the fermion frequency $\omega$:
\bea
m & = & \lambda \, m_u; \\
\omega & = & \lambda \,\tau\, \omega_u.
\eea
They obey the correct normalization condition,
\be
4 \pi \int\limits_0^\infty \left( \alpha^2 + \beta^2 \right) \frac{T}{\sqrt{A}} dr = 1,
\ee
and the metric field $T$ has the correct large-$r$ limit,
\be
\lim_{r \to \infty} T(r) = 1.
\ee

\begin{figure}
\begin{center}
\includegraphics[width=0.99\columnwidth]{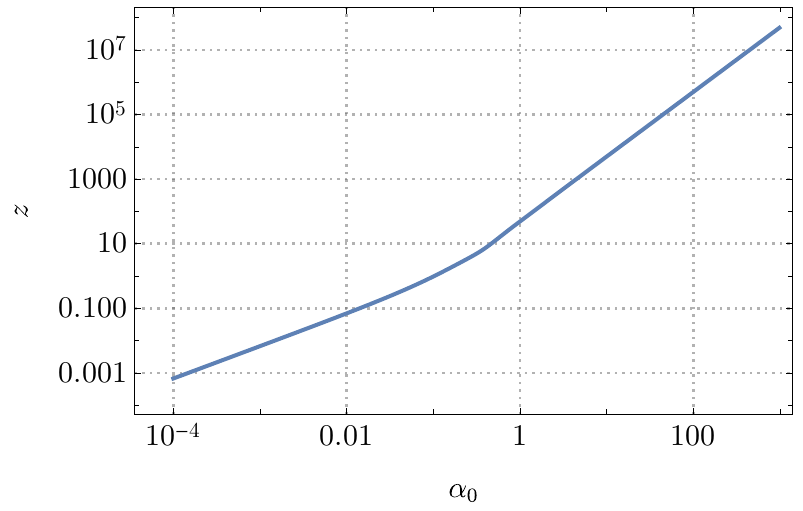}
\end{center}
\caption{One-to-one relationship between boundary condition parameter $\alpha_0$ and the central redshift $z=T(0)-1$ for ground state solutions with $N=2$ fermions. The change of gradient reflects the emergence of the new power law regime for the solutions.}
\label{fig:10}
\end{figure}

The parameter $\alpha_{0}$, which we chose arbitrarily at the beginning of the process, determines the central redshift $z=T(r=0)-1$ of the eventual scaled solution.  There is a one-to-one relationship between $\alpha_0$ and $z$ as shown in Figure \ref{fig:10}. All $\alpha_0$ values required to generate the solutions in this paper are presented, with their corresponding redshift, as follows: Figures 1-4 and 9  have $\alpha_0=20$ corresponding to a redshift $z=20,028$ for Figure 1 and $z= 18,247$ for Figures 2-4 and 9.

%\begin{table}[htp]
%\caption{default}
%\begin{center}
%\begin{tabular}{c|c|c}
%\quad Fig. \quad  &\quad $z$ \quad& \quad $\alpha_0$ \quad  \\
%\hline
%1 & e & e\\
%2 & e& e\\
%3 & e& e\\
%4 & e& e\\
%\end{tabular}
%\end{center}
%\label{default}
%\end{table}%

The procedure for numerically determining solutions of the FSY equations with infinite central redshift is similar, except that we use the power-law solution of the FSY equations from Section~\ref{s:purepower} in place of the small-$r$ expansion (\ref{smallru1})--(\ref{smallru4}).  The small-$r$ condition for the infinite-redshift solitons is thus given by
\bea
\alpha_{\rm PL}(r) & = & \alpha_{up} \, r, \label{plu1} \\
\beta_{\rm PL}(r) & = & \beta_{up} \, r, \\
A_{\rm PL}(r) & = & A_{up}, \\
T_{\rm PL}(r) & = & \frac{T_{up}}{r}. \label{plu4}
\eea
Here the `$u$' stands for `unscaled', while the `$p$' stands for `power-law'.  The coefficients in (\ref{plu1}--\ref{plu4}) are given by
\bea
	\alpha_{up}^2 & = &	
	\frac{ \omega_u } { 12 \, \pi \, G \, \sigma \, N^2 \, N_{u{-}} } 
\ ,
\hspace{5mm} 
	\left( \frac{ \beta_{up} } { \alpha_{up} } \right)^2
	= \frac{ N_{u{-}} } { N_{u{+}} } 
\ , \quad \nonumber \\
	A_{up} & = & \frac{ 1 } { 3 }
\ , 
\hspace{29mm} T_{up} =
	\frac{ \overline{ N }_u } { \omega_u }
\ ,
\eea
where $ \displaystyle N_{u\pm} \equiv \frac{ \sigma\, N} { 2 } \pm \sqrt{ A_{up} }$ and
\be
\overline{N}_u \equiv \left( N_{u{+}} \, N_{u{-}} \right)^{1/2}
	=  \left( \frac{ N^2}{4} - \frac{ 1 } { 3  } \right)^{1/2}
\ .
\ee
The rest of the procedure is as in the finite-redshift case:\ we tune $\omega_u$ until the fermion fields become normalizable, and then we use the same two-parameter scaling procedure to normalize the fermions and make the spacetime asymptotically flat as $r \to \infty$.

\section{Perturbations around the Power Law}
\label{appendixC}

To investigate the oscillations around the power-law solution
we consider perturbations of the power-law functions, introducing a small
parameter $\epsilon>0$:
\begin{align}
A(r) &= A_p \: (1 +\epsilon \: \overline{A}_1(r) + \mathcal{O}(\epsilon^2)) \\
T(r) &=  \frac{T_p}{r} \: (1 +\epsilon \: \overline{T}_1(r) + \mathcal{O}(\epsilon^2)  \\
\alpha(r) &= \alpha_p \,r \: (1 +\epsilon\: \overline{\alpha}_1(r) + \mathcal{O}(\epsilon^2)) \\
\beta(r) &=  \beta_p \, r \: (1 +\epsilon\: \overline{\beta}_1(r) + \mathcal{O}(\epsilon^2) )\\
%\phi(r) &= \phi_0
\end{align}
and linearizing the EDM equations for the behavior of the functions $\overline{A}_1$,  $\overline{T}_1$, 
$\overline{\alpha}_1$ and $\overline{\beta}_1$ 
(note that the overhead bar is mere notation and does not indicate conjugation). 
This reduces to a system of equations of the form
\be \frac{d\mathbf{y}}{d\ln{r}} = \mathbf{M} \;\mathbf{y} 
	+ m\, r \, \left( \mathbf{c} + \mathbf{S} \; \mathbf{y} \right)
\ ,
\ee
where 
$\mathbf{y}(r)=\left( 
	\overline{\alpha}_1(r), 
	\overline{\beta}(r),
	\overline{A}_1(r), 
	\overline{T}_1(r)
	\right)^T$
is the $1\times4$ column vector giving the $\mathcal{O}(\epsilon)$
perturbations around the power-law solution,
$\mathbf{c}$ is an $r$-independent $1\times4$ column vector,
$\mathbf{M}$ and $\mathbf{S}$ are $r$-independent $4\times4$ matrices.

Neglecting the $m$ terms, we have an eigenvalue problem with the matrix
\begin{align}
\mathbf{M} = \left(
\begin{array}{cccc}
 K-1 & 1-K & \: -\frac{1}{2}  \:& 1- K \\[5pt]
 1+K & -1-K & \: -\frac{1}{2} \:& 1+K \\[5pt]
 -2 - \frac{2}{K} & -2 +\frac{2}{K} & \: -1 \: & -4 \\[5pt]
 -1/K & 1/K & \: \frac{3}{2} \: & -1 \\
\end{array}
\right)
\end{align}
where $K\equiv\sqrt{3}\, \sigma\,N/2=(N_+ + N_-)/(N_+ - N_-)$.
The matrix $\mathbf{M}$ is independent of $r$ and depends only on the parameter $\sigma\,N$.  
The eigenvalue problem has solutions of the form
\begin{align} \label{eqn:gensoln}
	\mathbf{y}(r) = \sum_{j=1}^4 A_j \, \exp\left(\lambda_j \ln r\right) \: \mathbf{V}_j
	= \sum_{j=1}^4 A_j \, r^{\lambda_j}  \: \mathbf{V}_j
\ ,
\end{align}
where $\lambda_j$ are the eigenvalues and the corresponding right eigenvectors $\mathbf{V}_j$
are weighted by (complex) amplitudes $A_j$.

The trace ${\rm Tr}\left( \mathbf{M} \right)=-4$ is the sum of the 4 eigenvalues.
Writing the eigenvalues as $\lambda=q-1$, the constituent equation is
\be
	0 = {\rm Det}\left( \mathbf{M} - \lambda\, \mathbf{I} \right) 
	= q^4 + q^2 + 10 -12\, K^2 \ .
\ee
Note that $12\,K^2=9\,N^2$, so that the eigenvalues depend on $N$ but not on $\sigma$.
Solving the quadratic equation for $q^2$, the 4 roots are
\be
	q  = \pm \sqrt{ - \frac{1}{2} \pm\, \sqrt{ \frac{1}{4} + 12 \, K^2 - 10 } }
\ .
\ee
This gives 2 real eigenvalues
\be
	\lambda_{\pm p} = -1 \pm \sqrt{ - \frac{1}{2} + 3 \, \bar{N} } \equiv -1 \pm  p \ ,
\ee
and a complex conjugate pair
\be
	\lambda_{\pm k} = -1 \pm \sqrt{ -\frac{1}{2} - 3\, \bar{N} } \equiv -1 \pm i \, k \ ,
\ee
where
\be
	\bar{N} \equiv \sqrt{ N^2 - \frac{13}{12} } \ .
\ee
The two real eigenvalues, $\lambda_{\pm p} = -1 \pm p$, give one rising mode, $r^{p-1}$,
and one falling mode, $r^{-(1+p)}$.
The complex conjugate pair, $\lambda_{\pm k}=-1\pm i\,k$, combine to 
give damped oscillations in $\ln{r}$ with wavenumber $k$ and a $1/r$ decay envelope.
It is these damped oscillations that give rise to the spiral curves such as in Fig.~\ref{fig:7}.
The even spacing of the oscillations in $\ln{r}$ can be seen in log-log plots of the 
our numerical solutions, and the corresponding oscillations of 
$\omega$\, $m$, and $M$ with central redshift $\log{z}$ in Fig.~\ref{fig:8}.

The eigenvectors give the specific mix of fields involved in each of the four modes.
Solving $\mathbf{M} \, \mathbf{V} = \left(q-1\right)\, \mathbf{V}$,
the right eigenvectors of $\mathbf{M}$ are
%\be
%	\mathbf{V} \propto
%	\left( 
%	\ba{c}
%	-6\, K -q \, \left( q^2 + q + 4 \right) 
%	\\ - 6 \, K + q \, \left( q^2 + q + 4 \right)
%	\\ 4 \,  \left( q-1 \right) \, \left( q+2 \right)
%	\\ 6 \, \left( q + 1 \right)
%	\ea
%	\right) \ .
%\ee
\be
	\mathbf{V} \propto
	\left( 
	\ba{c}
	-6\, K -q^2 
	\\ - 6 \, K + q^2 
	\\ 4 \,  \left( q^2 - 2 \right)
	\\ 6 
	\ea
	\right) 
	+ q \left( \ba{c}
	-\left( q^2 + 4 \right)
	\\ + \left( q^2 + 4 \right)
	\\ 4
	\\ 6
	\ea \right)
\ .
\ee
For the two real eigenvalues,  $q=\pm p$,
 \be
	\mathbf{V}_{\pm p} \propto
	\left( 
	\ba{c}
	 - 6 \, K - p^2
	\\  - 6 \, K+ p^2
	\\ 4 \,  \left( p^2 - 2 \right)
	\\ 6 
	\ea
	\right) 
	\pm p \left( \ba{c}
	-\left( p^2 + 4 \right)
	\\ + \left( p^2 + 4 \right)
	\\ 4
	\\ 6
	\ea \right)
\ .
\ee
For the complex conjugate pair, $q=\pm i \, k$,
\be
	\mathbf{V}_{\pm k}  \propto
	\left( 
	\ba{c}
	-6\, K + k^2 
	\\ - 6 \, K - k^2 
	\\ -4 \,  \left( k^2 + 2 \right)
	\\ 6 
	\ea
	\right) 
	\pm i \, k \left( \ba{c}
	4 + k^2
	\\  4 - k^2 
	\\ 4
	\\ 6
	\ea \right)
\ .
\ee
In these expressions the eigenvectors are not normalized in any particular way.
The general solution given by (\ref{eqn:gensoln}) scales these 4 eigenvectors
by 4  complex amplitudes $A_j$.
Requiring the 4 field perturbations to be real provides 4 constraints,
leaving 4 real parameters. 
 The general solution, in terms of real functions of $r$, is thus
 \be \ba{rl}
 \mathbf{y}(r)
 	= & \displaystyle
	 A_+ \, \mathbf{V}_{+p} \, r^{p-1}
	+ A_C \, \mathbf{V}_C \, \frac{ \cos{ \left( k \, \ln{ r  } \right) } } { r }
 \\ \\ + &	\displaystyle
	A_- \, \mathbf{V}_{-p} \, r^{-1-p}
	+ A_S \, \mathbf{V}_S \, \frac{ \sin{ \left( k \, \ln{ r  } \right) } } { r } 
\ . \ea
 \ee
 Here 
$A_+$ and $A_-$ are (real) amplitudes for the rising and falling modes,
$A_S$ and $A_C$ are (real) amplitudes for the sine and cosine components of the damped oscillations,
and $\mathbf{V}_{\pm k}=\mathbf{V}_C \pm i \, \mathbf{V}_S$.

It is worth noting implications of the positive eigenvalue, $\lambda_{+p}$, which
causes perturbations to grow as $r^{p-1}$. 
The existence of this positive eigenvalue is not yet fully understood since the magnitude of its corresponding eigenvector does not seem to be negligible and yet the observed numerical solutions display damping of oscillations over lengths in $\ln{r}$ of order $\ln{z}$ for $z>>1$.
We conjecture that the fine-tuning of $\omega$ effectively suppresses the excitation of
this growing eigenmode so that its amplitude becomes important only at $r\sim 1$
where $\omega\,T<m$ marks entry into the evanescent zone.

The radial dynamics of the finite-redshift solutions may thus be summarized as follows:
The fields are launched from $r=0$ with the boundary conditions
$A=1$, $T=T_0=1+z$, $\alpha=\beta=0$.
In the core, $r<r_0$, the dominant $\sigma\,N/r$ terms in the Dirac equation
drive up the fermion amplitude as $r^{N/2}$.
With the fermion density then rising as $\eta\propto r^{N-2}$,
the metric fields $A$ and $T$ begin to decrease with an $r^N$ term.
In the high-redshift case, where $\omega\,T_0>>m$,
the fields approach the power-law solution, arriving at $r\sim r_0\approx 1/(1+z)$.
But they overshoot the power-law solution,
exciting the damped oscillations, and thus spiral around and converge 
toward the power-law solution until being ejected by the growing mode.
In the excited states, this ejection occurs where $\omega\,T(r)>m$, 
driving wave-zone oscillations until entry into the evanescent zone.
Once $\omega\,T(r)<m$,
the mass terms drive exponential decay of the fermions,
and leaving an exterior Schwarzshild metric.

\end{document}